\documentclass[prd, twocolumn, nofootinbib, floatfix]{revtex4-1}
\usepackage{amsmath}
\usepackage{graphicx}
\usepackage{dcolumn}
\usepackage{bm}
\usepackage{array}
\usepackage{epsfig}
\usepackage{amssymb,latexsym,mathrsfs}
\usepackage{graphicx}
\usepackage{color}

\newcommand{\be}{\begin{equation}}
\newcommand{\ee}{\end{equation}}
\newcommand{\bs}{\begin{split}}
\newcommand{\esplit}{\end{split}}
\newcommand{\bea}{\begin{eqnarray}}
\newcommand{\eea}{\end{eqnarray}}

\begin{document}
\title{From Asymptotic Safety to Dark Energy} 
\author{Changrim Ahn$^1$, Chanju Kim$^1$, Eric V.\ Linder$^{1,2}$}
\affiliation{$^1$ Institute for the Early Universe WCU and Department of 
Physics, Ewha Womans University, Seoul, Korea} 
\affiliation{$^2$ Berkeley Lab and University of California, Berkeley, USA} 

\date{\today}

\begin{abstract} 
We consider renormalization group flow applied to the cosmological 
dynamical equations.  A consistency condition arising from energy-momentum 
conservation links the flow parameters to the cosmological evolution, 
restricting possible behaviors.  Three classes of cosmological fixed 
points for dark energy plus a barotropic fluid are found: a dark energy 
dominated universe, which can be either accelerating or decelerating 
depending on the RG flow parameters, a barotropic dominated universe 
where dark energy fades away, and solutions where the gravitational and 
potential couplings cease to flow.   If the IR fixed point coincides with 
the asymptotically safe UV fixed point then the dark energy pressure 
vanishes in the first class, while (only) in the de Sitter limit of the 
third class the RG cutoff scale becomes the Hubble scale. 
\end{abstract}

\maketitle

\section{Introduction} 

Cosmic acceleration may be due to a scalar field arising from high 
energy physics.  This creates two puzzles: if the physics is set at 
the Planck scale, or similarly high energy, why is the magnitude of the 
energy density, basically the amplitude of the scalar field potential, 
of order $(10^{-3}\,{\rm eV})^4$ today, and why doesn't the potential 
amplitude and shape receive strong corrections from couplings in the 
high energy universe? 

A useful and efficient way of analyzing quantum effects on the low energy 
scale physics is the renormalization group (RG) \cite{wilson}.  
An effective theory is obtained by integrating out the quantum 
fluctuations with higher energy scales than a certain cutoff scale.  
It contains a number of parameters that run along with the cutoff 
scale, called the RG flows.  One can then incorporate the quantum effects 
using classical equations of motion from the effective action.  The main 
problem in applying the RG approach to cosmology is that we do not 
know the complete quantum gravity theory that governs the UV (Planck) scale 
physics.  Asymptotically safe gravity \cite{weinberg} is an idea that the 
quantum gravity is described by a finite number of parameters which 
approach non-trivial fixed points in the UV scale limit.  This provides a 
conceptual framework to link the UV physics with the low energy effective 
theory that describes physics at much later time scale. 

We explore here the cosmological late time effects from 
renormalization group flow.  This differs from the application of 
asymptotic safety criteria in the UV (see \cite{lrrniedermaier} for 
a review) in that we focus on the IR behavior of the field and its 
effects on dark energy and the cosmological expansion.  We look for 
cosmological fixed points to the coupled dynamical equations including 
RG effects, which may or may not correspond to fixed points of the RG flow. 

The low energy effective action can, in principle, be obtained from the 
RG equation.  This is however a highly nontrivial functional differential 
equation with respect to the RG scale $k$ \cite{wett93} that is virtually 
impossible to solve exactly. As a simple approximation, we (in agreement 
with much of the literature) shall adopt the Einstein-Hilbert truncation 
\cite{reut98} in the gravity part by neglecting higher derivative terms. 
In the matter part, the kinetic term of the scalar field is taken to be 
canonical (i.e.\ no running since there is no coupling parameter) while 
the potential is allowed to vary as the RG scale $k$ changes. 
In this approximation, the practical effect of RG flow is then
an evolution of the gravitational coupling $G_k$ (generalizing Newton's 
constant) and the scalar field potential $V_k(\phi)$. In particular the 
equations of motion will take the same form as the classical ones.

In the application of the RG to cosmology, the RG evolution governed by 
the RG cutoff $k$ is then related to the cosmological evolution in
time $t$. In the literature the cutoff $k$ is usually assumed to be
proportional to $1/t$ on the physical ground that fluctuations smaller 
than $1/t$ do not play any role, thus providing the IR cutoff 
\cite{bonanno,reut05}.  Another choice would be for example
$k \sim H(t)$, the Hubble parameter at $t$.  Note that $H\sim 1/t$ in 
general so this is quite similar.  See \cite{09061113,0609083} and 
references therein for this and other cutoffs, some inspired by 
holography, applied to cosmology. 

However, the truncation of the low energy effective action to the 
Einstein-Hilbert form already restricts the type of influence of the 
cutoff scale on the cosmology.  We will see that a constraint among RG 
parameters emerges for consistency of the approximation. The constraint 
has been discussed in previous works with a perfect fluid 
\cite{bonanno,reut05}.  In this paper we will consider the gravity with 
barotropic fluids and a scalar field and derive the constraint 
and its consequences.

In deriving the modifications of the Einstein field equations from the 
Einstein-Hilbert action with non-constant couplings, three approaches can 
be taken, depending on the interpretation used.  One method \cite{reut03} 
is to treat the evolution of the couplings as due to a dynamical variable, 
say a field $\phi$.
This is basically equivalent to a treatment like $F(\phi)\,R$ in the 
case of gravitational coupling, and leads to an extended quintessence 
type of scalar-tensor theory \cite{bacci,eq2,eq3,eq4,eq5}, with 
similarities to induced gravity \cite{induce1,induce2,induce3}. 

A second method is to keep the couplings as nondynamical during the 
variation of the action with respect to the metric, and further 
assume that continuity equation for each energy-momentum component 
individually is 
unaffected.  That is, require the part of the covariant derivatives with 
respect to spacetime coordinates and that part coming from a partial 
derivative of the renormalization scale $k$ with respect to spacetime 
coordinates to vanish separately.  This was the approach recently 
taken by \cite{hlr}.  Third, one can keep the couplings nondynamical 
in the action and require only the Bianchi identity to hold with respect 
to the total covariant derivatives, simultaneously accounting for the 
spacetime dependence and the flow of the couplings under the 
renormalization group.  This is the approach we take here, and its 
results, for example that the flow converges in de Sitter space, 
indicate that it is of interest in its physical consequences. 

In Sec.~\ref{sec:cos} we derive the effective dark energy contributions 
to the Friedmann equations and continuity equations, and the necessary 
consistency condition between them.  We evaluate the system of dynamical 
equations in Sec.~\ref{sec:dyn}, finding the cosmological fixed points. 
The relation of the RG cutoff scale to cosmology is addressed in 
Sec.~\ref{sec:cut} and we conclude in Sec.~\ref{sec:concl}.

\section{Cosmology with RG Flow} \label{sec:cos} 

We assume that the universe is described by Einstein gravity with matter 
(or other barotropic fluids) and a canonical, minimally coupled scalar field. 
In order to incorporate the quantum effects, we consider the truncated
RG flow leading to the Einstein-Hilbert action as explained above.  The 
couplings, including the gravitational coupling and the scalar field 
potential, will be assumed to run with scale.  Because the field equations 
arise from variation of the action with respect to the metric, and there 
is no explicit dependence of the couplings on the metric, the form of 
the field equations will be unaltered.  In particular, for a homogeneous 
and isotropic universe the standard form of the Friedmann equations for 
the expansion rate $H$ and the acceleration $\ddot a$ (or $\dot H$) will 
be preserved.  

The evolution equations are 
\bea 
H^2&=&\frac{8\pi G_k}{3}\rho_k \label{eq:fried1} \\ 
\dot H&=&-4\pi G_k\,(\rho_k+p_k) \,, \label{eq:fried2} 
\eea 
where $H=\dot a/a$, an overdot represents a time derivative, $\rho_k$ 
represents the total energy density including all components (e.g.\ 
matter, scalar field, etc.), and $p_k$ is the total pressure.  We 
show subscripts $k$ on the gravitational coupling (generalization of 
Newton's constant) $G_k$ and quantities involving the scalar field 
coupling, i.e.\ the potential, to remind that these may flow with 
the RG scale $k$. 

Note that in a scalar-tensor theory, where the time variation of the 
gravitational coupling arises from a dynamical field, the form of the 
Friedmann equations {\it will\/} be modified.  Extra terms involving 
$\dot G$ and $\ddot G$ will appear. 

One also has the Bianchi identity, involving the covariant derivative 
of each side of the Einstein field equation.  This gives 
\be 
\bs 
0&=\left(G_k\,T_k^{\mu\nu}\right)_{;\nu}\\ 
&=(G_k\,T_k^{\mu\nu})_{,\nu} + G_k\,\Gamma^\mu{}_{\alpha\nu}T_k^{\alpha\nu} 
+ G_k\,\Gamma^\nu{}_{\alpha\nu} T_k^{\mu\alpha} \,. 
\end{split} 
\ee 
For the $\mu=0$ equation in a Friedmann-Robertson-Walker cosmology one gets 
\be 
(G_k \rho_k)_{,0}+3HG_k(\rho+p_k)=0\,. 
\ee 

Finally, one must take into account that the time derivative involves 
a piece from the possible time variation of the RG scale $k$ to find 
the continuity equation 
\be 
\frac{\partial\rho_k}{\partial t}=-3H\rho_k\left[1+\frac{p_k}{\rho_k}+ 
\frac{1}{3}\frac{d\ln k}{dN}\frac{\partial\ln(G_k\rho_k)}{\partial\ln k}\right]\,, 
\ee 
where $N=\ln a$.  
Checking against $\partial\rho_k/\partial t$ derived by differentiating 
Eq.~(\ref{eq:fried1}) and substituting into Eq.~(\ref{eq:fried2}) gives 
agreement.  That is, preservation of the form of the Friedmann equations 
necessarily implies a modification to the continuity equation due to 
the flow of the RG scale. 

For the matter component, the continuity equation reads 
\be 
\partial_t \rho_m=-3H\rho_m\, 
\left[1+\frac{1}{3}\frac{\partial\ln G_k}{\partial\ln k}\frac{d\ln k}{dN}\right]\,.
\ee 
Note that the evolution is altered from the usual behavior 
due to the flow of the gravitational coupling.  For the dark energy 
component, the evolution is 
\be 
\bs 
\partial_t\rho_{de}&=-3H(\rho_{de}+p_{de})\\ 
&\quad -H\,\frac{d\ln k}{dN} 
\left(\frac{\partial\ln G_k}{\partial\ln k}\,\rho_{de}+\frac{\partial\ln V_k}{\partial\ln k}\,V_k\right)\,. \label{eq:dotde} 
\end{split} 
\ee 
Note that only the coupling coefficients within the potential change under 
the RG flow, and the kinetic term and matter density are unchanged.  
For the total density, the continuity equation is 
\be 
\bs 
\partial_t \rho_k&=-3H(\rho_k+p_k)\\ 
&\quad -H\frac{d\ln k}{dN} 
\left(\frac{\partial\ln G_k}{\partial\ln k}\,\rho_k+\frac{\partial\ln V_k}{\partial\ln k}\,V_k\right)\,. 
\end{split} 
\ee 
We can verify that the total density equation is indeed consistent 
with the sum of the individual components, as another check on the 
system of equations. 

However, a crucial further condition is that the variation of the action 
with respect to the field $\phi$ gives an unaltered Klein-Gordon equation, 
since there is no explicit $k$ dependence of $\phi$.  This field equation 
must be consistent with the continuity equation we derived.  Introducing 
the RG flow parameters (also called the anomalous dimensions) arising from 
the flow of the effective action, 
\bea 
\eta&\equiv&\frac{\partial\ln G_k}{\partial\ln k} \\ 
\nu&\equiv&\frac{\partial\ln V_k}{\partial\ln k} \,, 
\eea 
and taking the derivative of $\rho_{de}=(1/2)\dot\phi^2+V_k$ we find 
\be 
\begin{split} 
\partial_t\rho_{de}&=\dot\phi\ddot\phi 
+\frac{\partial V}{\partial\phi}\dot\phi\\ 
&=-3H\dot\phi^2-H\eta\frac{d\ln k}{dN}\left(\frac{1}{2}\dot\phi^2+V_k\right) 
-H\nu\frac{d\ln k}{dN}V_k\,, \label{eq:kg} 
\end{split} 
\ee 
where the second line comes from the continuity equation (\ref{eq:dotde}) 
for the dark energy.  Only the first term of the second line appears in 
the Klein-Gordon equation, though, so consistency of the theory requires 
that the terms proportional to $d\ln k/dN$ must vanish. 

This condition leads to two possibilities: either $d\ln k/dN=0$ for all time, 
in which case there is no relation between cosmological evolution and 
renormalization group flow, or 
\be 
0=\frac{1}{2}\dot\phi^2 \eta+V_k\,(\eta+\nu)\,. \label{eq:consphi}
\ee 
This is a crucial point because it restricts arbitrary behavior of the 
RG flow cosmology for the truncated, i.e.\ Einstein-Hilbert action.  

In summary, the gravitational field equations, Bianchi identity, and field 
evolution equation give a consistent framework within which to treat 
renormalization group flow and cosmological dynamics together when 
Eq.~(\ref{eq:consphi}) is applied.

\section{System of Dynamical Equations} \label{sec:dyn} 

To evaluate the cosmological evolution we can rewrite the cosmological 
evolution equations in the standard way (see, e.g., \cite{copelw}) as a 
coupled system of equations for the dynamics.  We use the dynamical variables 
\bea 
x^2&=&\frac{\kappa^2\dot\phi^2}{6H^2} \\
y^2&=&\frac{\kappa^2 V_k}{3H^2} \,,
\eea 
where $\kappa^2=8\pi G_k$.  The system of equations is 
\bea 
\frac{dx}{dN}&=&-3x(1-x^2)+\sqrt{\frac{3}{2}}\lambda y^2+\frac{3}{2}\Sigma x
+\frac{\eta}{2} x\frac{d\ln k}{dN} \\
\frac{dy}{dN}&=&-\sqrt{\frac{3}{2}}\lambda xy+3x^2 y +\frac{3}{2}\Sigma y
+\frac{\eta+\nu}{2} y\frac{d\ln k}{dN} \,,
\eea 
where the logarithmic potential slope
\be
\lambda\equiv -\frac{1}{\kappa V_k}\frac{dV_k}{d\phi} \,, \label{eq:lamb} 
\ee 
and $\Sigma=\sum_{i\ne de} (1+w_i)\Omega_i(a)$.
The sum includes all barotropic fluids present such as matter or
radiation, but the scalar field component is treated separately.  Here
$\Omega_i$ is the dimensionless energy density in barotropic component
$i$ and $w_i$ is its equation of state parameter or pressure to density 
ratio (e.g.\ 0 for matter, $1/3$ for radiation).

Thus the time dependence of the RG cutoff parameter $k(N)$ will be an
important element in the cosmological dynamics. 

Cosmological quantities of interest will be the effective dark energy
density and its equation of state defined through $d\ln\rho_{de}/dN=-3(1+w)$, 
\bea 
\Omega_{de}&=&x^2+y^2 \\
w&=&\frac{x^2-y^2}{x^2+y^2} \,. 
\eea 
Note that the influence of the RG flow is implicit in the behavior 
of $x$ and $y$; the correction term in Eq.~(\ref{eq:dotde}) 
vanishes due to the consistency condition and so the scale $k$ does not 
explicitly appear. 

In the case where $d\ln k/dN=0$ for all time, there is no RG flow and 
the standard cosmological dynamics results apply.  We therefore do not 
consider this case further.  The necessary consistency condition of 
Eq.~(\ref{eq:consphi}) then becomes in terms of the dynamical variables, 
\be 
y^2=\frac{-\eta}{\eta+\nu}\,x^2\,. \label{eq:cons} 
\ee 
(Note that we expect $\eta$ to be negative.)  Applying this to the 
cosmological quantities gives 
\bea 
\Omega_{de}&=&x^2\,\frac{\nu}{\eta+\nu} \\
w&=&\frac{2\eta+\nu}{\nu} \,. 
\eea 

We will be particularly interested in fixed point solutions of the 
dynamics, asymptotic behaviors that are insensitive to initial conditions 
and can serve as cosmological, and possibly RG flow, attractors. 

In searching for such solutions, first consider $y=0$.  Then 
the solutions are either $x=0$, which implies $\Omega_{de}=0$, i.e.\ the 
vanishing of dark energy, or $d\ln k/dN=0$, $x=1$ and $\Sigma=0$, 
i.e.\ complete dark 
energy domination with $\Omega_{de}=1$ but $w=1$ so this is a kinetic 
energy dominated solution that decelerates the expansion.  Both of these 
are standard cosmology solutions in the absence of RG flow, since 
asymptotically $d\ln k/dN=0$, i.e.\ the RG flow freezes.  However, the 
trajectory to reach the fixed point in general differs in the RG cosmology. 

If $x=0$ the fixed points are $y=0$ as already considered, or $d\ln k/dN=0$ 
with $\lambda=0$ (as in a runaway, e.g.\ inverse power law potential).  
This solution is dark energy dominated with $\Omega_{de}=1$ 
and $w=-1$, so this is a potential energy dominated case that accelerates 
the expansion, ending in a de Sitter state.  Again, this asymptotically 
agrees with a standard cosmology fixed point.  

In the case where asymptotically $d\ln k/dN=0$ (but $x\ne0\ne y$), the 
critical points are 
\begin{align}
x^2_{c1}&=\frac{\lambda^2}{6} & ;\qquad  x^2_{c2}&=\frac{3}{2}
\frac{(1+w_b)^2}{\lambda^2} \notag \\
y^2_{c1}&=1-\frac{\lambda^2}{6} & ;\qquad  y^2_{c2}&=\frac{3}{2}
\frac{1-w_b^2}{\lambda^2} \notag \\
\Omega_{de,c1}&=1 & ;\! \quad \Omega_{de,c1}&=\frac{3(1+w_b)}{\lambda^2} \notag \\
w_{c1}&=-1+\frac{\lambda^2}{3} & ;\qquad  w_{c2}&=w_b 
\end{align} 
The first critical point is dark energy dominated, with an equation of state 
depending on the value of $\lambda$.  If $\lambda=0$ asymptotically, then 
the dynamics approaches a de Sitter state.  A stable fixed point only 
exists for $\lambda^2<3$.  The second critical point is a scaling solution 
where dark energy and the least positive equation of state barotropic 
component have densities in a constant ratio.  Since $w$ is equal to the 
equation of state of the barotropic component $w_b$ then this cannot give 
acceleration unless one already had an accelerating barotropic component. 

Going beyond these cases, there is only one general solution since $y$ 
is not independent of $x$ due to the consistency requirement of 
Eq.~(\ref{eq:cons}), 
\be 
\bs 
x^2_{c3}&=1+\frac{\eta}{\nu}\\ 
y^2_{c3}&=\frac{-\eta}{\nu}\\ 
\Omega_{de,c3}&=1\\ 
w_{c3}&=1+\frac{2\eta}{\nu} 
\end{split} 
\ee 
This is a dark energy dominated solution with the possibility of a 
variety of equations of state, depending on the specific renormalization 
group theory.  In particular, the case with $\eta=-2$, $\nu=4$ 
corresponding to the asymptotically safe UV fixed point for the RG flow 
(not the cosmology) gives $w=0$ asymptotically, i.e.\ the cosmological 
dynamics behaves asymptotically in the future like a matter dominated 
universe. 

The RG scale parameter here evolves at the cosmological fixed point as 
\be 
\frac{d\ln k}{dN}=\frac{2}{\nu} 
\left[-3+\lambda\sqrt{\frac{3}{2}\frac{\nu}{\eta+\nu}}\right] \,. 
\ee 
Recall that the matter in general does not have an equation of state of 
zero, but rather 
\be 
w_m=\frac{\eta}{3}\frac{d\ln k}{dN} \,. 
\ee 
If one wanted $w_m=0$ then one must shut off the evolution of the 
scale $k$ through choosing $\lambda^2=6(1+\eta/\nu)$.  
However, the matter is irrelevant asymptotically in the dark energy 
dominated solution above.  (Also see the next section for further 
discussion.) 

\begin{figure}[htbp!]
\includegraphics[width=\columnwidth]{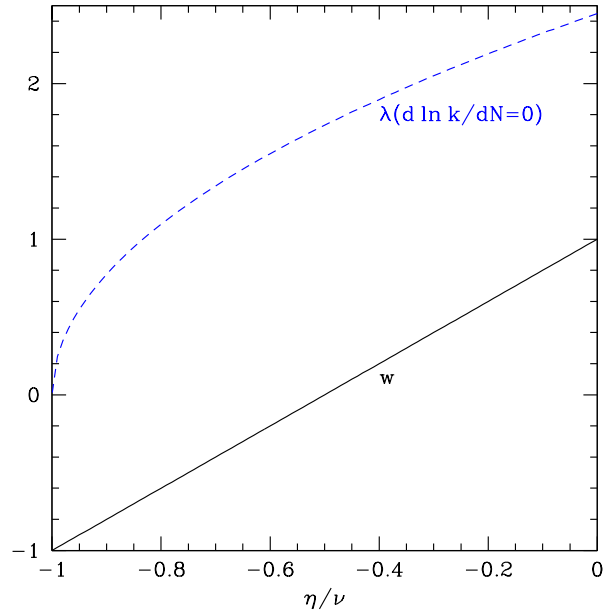}
\caption{The ratio $\eta/\nu$ of the anomalous dimensions of the RG 
flow variables determines the cosmological fixed point for the third 
critical point.  The solid 
black curve shows the dark energy equation of state dominating the 
future cosmic expansion, while the blue dashed curve shows the value 
of $\lambda$ for which the RG flow freezes.  Runaway potentials such 
as inverse power laws give $\lambda=0$ and so $w=-1$. 
}
\label{fig:runnv}
\end{figure}

Figure~\ref{fig:runnv} shows the dependence on $\eta/\nu$ of the fixed 
point values for the dark energy equation of state $w$ in this case, and 
the potential slope $\lambda$ needed to freeze the RG flow.

In order to end in a de Sitter state in this case, one needs $\eta=-\nu$.  
Going back to the original consistency condition on the Klein-Gordon equation 
(\ref{eq:kg}), this requires $\eta x^2\,(d\ln k/dN)=0$.  So either we 
reduce to the previous $d\ln k/dN=0$ solution that gave de Sitter 
behavior, or we take $\eta=0=\nu$, which requires the previous $x=0$, 
$\lambda=0$ solution.  Thus the list of cases giving $w=-1$ is complete. 

The cosmological attractor solutions are summarized in Table~\ref{tab:attsol}. 
The asymptotic de Sitter solutions with $\lambda=0$ can be achieved by a 
runaway potential such as an inverse power law $V\sim\phi^{-n}$ 
\cite{ratrap}, with the field $\phi$ rolling to the zero potential 
minimum, without the need for an explicit cosmological constant.

\begin{table*}[htbp]
\begin{center}
\renewcommand{\arraystretch}{1.5} 
\begin{tabular*}{2.\columnwidth}
{@{\extracolsep{\fill}} c c c c c c c}
\hline 
$x^2$ & $y^2$ & $\lambda$ & $d\ln k/dN$ & $w$ & $\Omega_{de}$ & Type \\
\hline 
$\frac{\lambda^2}{6}$ & $1-\frac{\lambda^2}{6}$ & $\lambda$ & 0 & 
$-1+\frac{\lambda^2}{3}$ & $1$ & Accelerating DE dominated\\ 
$\frac{3}{2}\frac{(1+w_b)^2}{\lambda^2}$ & $\frac{3(1-w_b^2)}{2} 
\frac{1}{\lambda^2}$ & $\lambda$ & 0 & $w_b$ & $\frac{3(1+w_b)}{\lambda^2}$ 
& Scaling\\ 
$1+\frac{\eta}{\nu}$ & $\frac{-\eta}{\nu}$ & $\lambda$ & $\frac{-6}{\nu} 
+\frac{\lambda}{\nu}\sqrt{\frac{6\nu}{\eta+\nu}}$ & $1+\frac{2\eta}{\nu}$ 
& 1 & Flowing DE dominated\\ 
0 & 0 & $\lambda$ & $d\ln k/dN$ & $w$ & 0 & Barotropic dominated\\ 
0 & 1 & 0 & 0 & $-1$ & 1 & DE potential dominated\\ 
1 & 0 & $\lambda$ & 0 & 1 & 1 & DE kinetic dominated\\ 
0 & 1 & 0 & $d\ln k/dN$ but $\eta=0=\nu$ & $-1$ & 1 & no RG, de Sitter\\ 
\hline
\end{tabular*}
\end{center}
\caption{Cosmological attractor solutions under the renormalization group 
flow, including the values of the dark energy equation of state $w$, energy 
density $\Omega_{de}$, and type of solution.  When a variable is repeated 
under its column heading that means its value is moot. 
} 
\label{tab:attsol}
\end{table*}

\section{Relation of Cutoff Scale to Hubble Scale} \label{sec:cut} 

Finally, let us  examine the issue of the dependence of the 
renormalization cutoff scale $k$ on the Hubble scale $H$.  Taking the 
derivative of Eq.~(\ref{eq:fried1}) with respect to $\ln k$ one obtains 
(also see \cite{hlr} for the first equality) 
\be 
\frac{\partial\ln H^2}{\partial\ln k}=\eta+\nu y^2= 
\eta\,[1-\Omega_{de}(t)] \,, 
\ee 
where the second equality follows from our consistency condition. 

In general the right hand side evolves with time so an explicit dependence 
of $k$ on $H$, such as $k\sim H^p$ which gives a constant left hand side, 
without time explicitly entering, would 
be very special.  Such a relation, which is sometimes assumed in the RG 
cosmology literature, will not in general be consistent. 

A special case is when $\Omega_{de}(t)={\rm constant}$ is achieved through 
$\lambda={\rm constant}$ for all time, i.e.\ an exponential potential 
\cite{wett88}.  This situation implies that dark energy is either the only 
component if $\lambda^2<3$, or scales with the barotropic component 
otherwise; such a universe does not yield acceleration.  Together with 
this must go that $\eta$ and $\nu$ are constant.  Thus, the assumption of 
the RG cutoff scale $k$ being proportional to the Hubble scale, or some 
power of it, is extremely restricting, and does not lead to viable 
solutions describing our universe. 

If we want to know how $k$ asymptotically depends on $H$ at the 
cosmological fixed point, we see that it can there have a power law 
relation with $H$.  For example, in the dark energy potential dominated 
solution one gets asymptotically $k\sim H^{2/(\eta+\nu)}$.  If one 
wanted the IR fixed point to return to the asymptotically safe UV fixed 
point of $\eta=-2$, $\nu=4$, this would give $k\sim H$ in the future 
limit (but not for the present or all times in general).  

In addition, note that astrophysical conditions exist on the flow of the 
gravitational 
coupling.  Observations of the cosmic microwave background \cite{zahn} 
and primordial nucleosynthesis abundances \cite{bbn} indicate that the 
coupling is constant to a precision of $\sim10\%$ over a time comparable 
to the age of the universe.  Thus, 
\be 
\frac{\dot G}{G}=\frac{1}{H}\frac{d\ln G}{dN}=\frac{1}{H}\frac{d\ln k}{dN} 
\,\eta<\frac{0.1}{H} \,, 
\ee 
places a condition $\eta\,(d\ln k/dN)<0.1$.  This can be achieved either 
through a small magnitude of $\eta$ or a slow flow $d\ln k/dN$ since 
primordial nucleosynthesis ($\sim 1\,$MeV scale).  The dashed curve in 
Figure~\ref{fig:runnv} shows the condition on $\lambda$ needed to give 
$d\ln k/dN=0$ for the flowing DE critical point, for example.  This will 
simultaneously also ensure that the matter equation of state $w_m=0$.

\section{Conclusions \label{sec:concl}} 

In this paper we have explored the quantum modifications to cosmological
evolution at late times.  For Einstein gravity and barotropic fluid and 
scalar field components, we considered the RG running of the gravitational
coupling (Newton's constant) and the scalar field potential.  Keeping the 
form of the equations of motion invariant under the RG evolution leads to 
a necessary consistency condition between the RG flow parameters and the 
cosmological quantities.  This condition implies that one cannot adopt 
an arbitrary a priori relation between the RG cutoff scale $k$ and the 
cosmological Hubble parameter $H(t)$.  

From the RG influenced cosmological evolution equations, we have identified 
three classes of cosmological fixed points depending on the RG parameters: 
a dark energy dominated universe, a barotropic dominated universe, and 
solutions where the gravitational and potential couplings cease to flow. 
One can obtain an asymptotically de Sitter universe with $w=-1$ for 
specific choices of parameters, even if the potential has no intrinsic 
cosmological constant. 

In general, due to the flow of the gravitational coupling the matter 
equation of state is not zero.  This will affect structure formation, 
which is beyond the scope of this article, but the requirements on the 
parameters are similar to those directly on variation of $G$.  We have 
considered cosmological constraints on $\dot G/G$ from cosmic microwave 
background and primordial nucleosynthesis observations and given the 
conditions necessary on the flow behavior.  One can also satisfy both 
the matter equation of state and varying gravity requirements through 
specific choices of potential. 

In this paper we have not specified an explicit form of the scalar field 
potential.  It would be interesting in future work to solve the RG equation 
explicitly for various specific potentials, such as those just mentioned, 
and see how the cosmological evolution develops toward the fixed points 
we have found.  One could also consider the higher order 
terms beyond the conventional truncation as used here and see how the 
consistency condition is modified.  This would serve as a test of the 
renormalization group formalism as usually applied to cosmology.

\acknowledgments

This work has been supported by the World Class University grant 
R32-2009-000-10130-0 through the National Research Foundation, Ministry 
of Education, Science and Technology of Korea.  EL is also supported in 
part by the Director, Office of Science, Office of High Energy Physics, 
of the U.S.\ Department of Energy under Contract No.\ DE-AC02-05CH11231.



\begin{thebibliography}{}

\bibitem{wilson}
K. G. Wilson and J. Kogut, Phys. Repts.  {\bf 12C}, 75 (1974)

\bibitem{weinberg}
S. Weinberg, Phys. Rev. D {\bf 81}, 083535 (2010)

\bibitem{lrrniedermaier} 
M. Niedermaier \& M. Reuter, Liv. Rev. Relativity {\bf 9}, 5 (2006) 

\bibitem{wett93}
C.~Wetterich, Phys.\ Lett.\  B {\bf 301}, 90 (1993)

\bibitem{reut98}
M.~Reuter, Phys.\ Rev.\  D {\bf 57}, 971 (1998)

\bibitem{bonanno}
A.~Bonanno and M.~Reuter, Phys.\ Rev.\  D {\bf 65}, 043508 (2002)

\bibitem{reut05}
M.~Reuter and F.~Saueressig, JCAP {\bf 0509}, 012 (2005)

\bibitem{09061113} 
L. Xu, Mod. Phys. Lett. A {\bf 25}, 377 (2010) 

\bibitem{0609083} 
J. Grande, J. Sola, H. Stefancic, Phys. Lett. B {\bf 645}, 236 (2007)

\bibitem{reut03}
M.~Reuter and H.~Weyer, Phys.\ Rev.\  D {\bf 69}, 104022 (2004)

\bibitem{bacci} 
F. Perrotta, C. Baccigalupi, and S. Matarrese, Phys. Rev. D 
{\bf 61}, 023507 (2000)

\bibitem{eq2} 
N. Bartolo and M. Pietroni, Phys. Rev. D {\bf 61}, 023518 (2000) 

\bibitem{eq3} 
C. Baccigalupi, S. Matarrese and F. Perrotta, Phys. Rev. D {\bf 62}, 
123510 (2000) 

\bibitem{eq4} 
T. Chiba, Phys. Rev. D {\bf 64}, 103503 (2001)

\bibitem{eq5} 
V. Faraoni and M.N. Jensen, Classical Quantum Gravity {\bf 23}, 
3005 (2006)

\bibitem{induce1} 
P. Minkowski, Phys. Lett. B {\bf 71}, 419 (1977) 

\bibitem{induce2} 
A. Zee, Phys. Rev. Lett.  {\bf 42}, 417 (1979) 

\bibitem{induce3} 
S.L. Adler, Rev. Mod. Phys. {\bf 54}, 729 (1982) 

\bibitem{hlr} 
M. Hindmarsh, D. Litim, C. Rahmede, arXiv:1101.5401

\bibitem{copelw} 
E.J.\ Copeland, A.R.\ Liddle, D.\ Wands, Phys.\ Rev.\ D {\bf 57}, 4686 (1998)

\bibitem{ratrap} 
B.\ Ratra \& P.J.E.\ Peebles, Phys. Rev. D {\bf 37}, 3406 (1988)

\bibitem{wett88} 
C.\ Wetterich, Nucl.\ Phys.\ B {\bf 302}, 668 (1988) 

\bibitem{zahn} 
S. Galli, A. Melchiorri, G.F. Smoot, O. Zahn, Phys. Rev. D {\bf 80}, 023508 
(2009) 

\bibitem{bbn} 
J-P. Uzan, Liv. Rev. Relativity {\bf 14}, 2 (2011)

\end{thebibliography}
\end{document}